\documentclass[twocolumn,times]{article}
\usepackage{amsmath}
\usepackage{xcolor}
\usepackage{color,soul}

\usepackage{underscore}

\usepackage{subfig}
\usepackage{graphicx}
\usepackage{amssymb}

\usepackage{setspace}

\usepackage{ragged2e}
\usepackage{blindtext, graphicx}
\graphicspath{{Pictures/}{Profile/}} 

\advance\topmargin-1in\advance\oddsidemargin-1in
\evensidemargin\oddsidemargin

\makeatletter
\def\@normalsize{\@setsize\normalsize{12pt}\xpt\@xpt
\abovedisplayskip 10pt plus2pt minus5pt\belowdisplayskip \abovedisplayskip
\abovedisplayshortskip \z@ plus3pt\belowdisplayshortskip 6pt plus3pt
minus3pt\let\@listi\@listI}

\def\section{\@startsection {section}{1}{\z@}{20pt plus 2pt minus 2pt}
{8pt plus 2pt minus 2pt}{\centering\normalsize\sc
\edef\@svsec{\thesection.\ }}}
\def\thesection{\Roman{section}}

\def\subsection{\@startsection {subsection}{2}{\z@}{16pt plus 2pt minus 2pt}
{6pt plus 2pt minus 2pt}{\normalsize\sl
\edef\@svsec{\thesubsection.\ }}}
\def\thesubsection{\Alph{subsection}}

\long\def\@makecaption#1#2{
\vskip10pt\begin{center} #1 #2 \end{center}\par\vskip 1pt}
\def\fnum@figure{\raggedright{\footnotesize Fig. \thefigure }.%
\footnotesize}
\def\fnum@table{\footnotesize TABLE \thetable\\\footnotesize\sc}
\def\thetable{\Roman{table}}

\makeatother

\begin{document}
\date{}
\title{\Large\bf
ROBIN: Inc\underline{r}emental \underline{Ob}lique \underline{In}terleaved ECC for Reliability Improvement in STT-MRAM Caches \\~\\
\large\bf
}\vspace{-50pt}



\author{
{\fontfamily{ptm}\selectfont
{Elham~Cheshmikhani}$^\ast$, Hamed Farbeh$^\dagger$, and Hossein Asadi$^\ast$}\\
{\fontfamily{ptm}\selectfont
$^\ast$Department of Computer Engineering, Sharif University of Technology, Tehran, Iran}  \\
{\fontfamily{ptm}\selectfont
$^\dagger$Department of Computer Engineering, Amirkabir University of Technology, Tehran, Iran}\\ 
{{\fontfamily{ptm}\selectfont$^\ast$\{elham.cheshmikhani, asadi\}@sharif.edu; $^\dagger$farbeh@aut.ac.ir}
}
}

\maketitle
\thispagestyle{empty}
\vspace{-10pt}
{\small\bf Abstract---
\textit{Spin-Transfer Torque Magnetic RAM} (STT-MRAM)  is a promising alternative for SRAMs in on-chip cache memories. Besides all its advantages, high error rate in STT-MRAM is a major limiting factor for on-chip cache memories.
In this paper, we first present a comprehensive analysis that reveals that the conventional \textit{Error-Correcting~Codes} (ECCs) lose their efficiency due to data-dependent error patterns, and then propose an efficient ECC configuration, so-called \textit{ROBIN}, to improve the correction capability.
The evaluations show that the inefficiency of conventional ECC increases the cache error rate by an average of 151.7\% while ROBIN reduces this value by more than 28.6x.}
\vspace{-10pt}

\section{Introduction}

\textit{Spin-Transfer Torque Magnetic RAM} (STT-MRAM) has attracted considerable research interests and efforts in recent years~\cite{naeimi2013intel,Eli-TC,salkhordeh2016operating}.
Non-volatility, near-zero leakage power, high density, and immunity to radiation-induced particle strike persuade the designers to replace conventional SRAM technology with STT-MRAM in \textit{Last-Level Caches} (LLCs)~\cite{Eli-TC, Chen2016, vatajelu2017challenges}.
Beside all of STT-MRAM technology advantages, it suffers from three error types: \textit{write~failure} (i.e., unsuccessful cell flip during a write operation), \textit{read disturbance} (i.e., unintentional cell flip during a read operation), and \textit{retention~failure} (i.e., stochastic cell flip during cell idle interval)~\cite{naeimi2013intel, Eli-TC, choi2017nvm, mittal2017survey, Chintaluri-ESTCS2016}.
To make STT-MRAM technology commercialized in on-chip caches, these reliability challenges should be  carefully addressed by designers.


Employing \textit{Error-Correcting~Codes} (ECCs) is the most conventional scheme to overcome write failure in STT-MRAM LLCs, which is the focus of this study~\cite{ZAZADTPDS,farbeh2016floating,ZAZADTE, guo2017sanitizer, 16-AZAD-TETC}. 
Conventionally, each \textit{N}-bit cache block is divided into several \textit{k}-bit datawords, each of which is protected by an \textit{r}-bit ECC. 
There are two common configurations for partitioning cache block cells to construct \textit{(k+r)}-bit codewords.
In \textit{per-word} ECC configuration, a codeword is generated by \textit{k} consecutive bit positions and in \textit{interleaved ECC} configuration, bit positions with distance of \textit{N/k} contribute in generating each codeword.

Since write failure is likely to occur \textit{only} in cache cells that need to flip in a write operation, the write failure rate is proportional to the number of transitions required.
When partitioning a block into multiple codewords, each capable of correcting single bit error, the probability of a correct write operation is dominated by a codeword with the maximum number of transitions. 
This number should be minimized for maximizing the ECC efficiency, which can be achieved by uniformly distributing the total number of transitions between the codewords. 
This uniformity is addressed in neither per-word nor interleaved ECCs. 
To improve the efficiency of ECCs and provide higher error correction capabilities, the ECC configuration should be redesigned based on STT-MRAM error patterns and customized according to its characteristics and requirements, which has been addressed in none of the previous studies.

In this paper, we first demonstrate the inefficiency of existing ECC configurations and then propose an efficient ECC configuration that effectively improve the reliability of STT-MRAM LLCs. In the first contribution, we conduct a deep investigation on the distribution of transitions between codewords of cache blocks in write operations and observe a large variations between the number of transitions in different codewords for both per-word and interleaved configurations, which cause a significant reliability degradation. In the second contribution, we propose ${Inc\underline{r}emental~\underline{Ob}lique~\underline{In}terleaved}$~ (ROBIN) ECC configuration to uniformly distribute the transitions between the codewords. ROBIN selects the data bits of each codeword in such a way that all bytes of a cache block as well as all bit positions in all bytes equally contribute in all codewords. Meanwhile, the bit positions of the bytes in each word are shifted in such a way that bits of different byte positions in all words uniformly contribute in all codewords.  
The uniformity achieved by ROBIN significantly improves the ECCs efficiency and  provides higher error correction capability.

We evaluate ROBIN using gem5 cycle-accurate simulator~\cite{gem5} and compare it with per-word ECC and interleaved ECC.
The evaluations show that the error rate in per-word and interleaved ECCs is higher than the error rate in optimal ECC by 151.7\% and 42.3\%, respectively, which is reduced to 5.3\% by ROBIN (equivalent to 28.6x and 8.0x improvements, respectively).
These significant improvements are achieved without increasing the overheads and complexity of ECCs.

The rest of this paper is organized as follows. Section II describes the preliminaries of STT-MRAM memory and write failure. 
In Section III, the motivation and the proposed method is presented.
In Section IV, the simulation framework and results are given.
Finally, we conclude the paper in Section V.
\vspace{-10pt}

\section{STT-MRAM Basics}

STT-MRAM memories consist of an access transistor and a \textit{Magnetic~Tunnel~Junction} (MTJ), which has three layers:
\textit{Reference~layer}, which its magnetization direction is fix, \textit{free~layer}, which its magnetization direction can be changed and determines the stored data, and the middle layer is \textit{oxide~barrier~layer}, which separates these two ferromagnetic layers.
The magnetization of free layer due to spin-polarized current flow can be in the same or opposite direction as reference layer spin direction.
This parallelism and anti-parallelism of two layers generates low and high resistance in MTJ that is interpreted as logic value `0' and `1' in the cell, respectively~\cite{asadi2017wipe}.

A data is written into a STT-MRAM cell by applying a write current (\textit{I$_{write}$}) for a predetermined pulse width.
Based on its direction, this current flow changes the magnetic field direction of the free layer and generates high or low resistance in the MTJ.
Changing this direction from parallel to anti-parallel (or vice-versa) is a stochastic process.
This means that by applying write current, it is probable that the cell remains unflipped. 
This unsuccessful write operation is named \textit{write failure}~\cite{ZAZADTPDS, ZAZADTE,14-zazad-eken2014novel}.
The probability of a write failure in a STT-MRAM cell is according to (\ref{eq:1}):\vspace{-5pt}
	\begin{flalign}
	\begin{split}
			\label{eq:1}
			P_{Write-Failure}= 1 - P_{write}= exp( -t_{write}\times   \\
			\frac{2 \times \mu_{\beta}\times p\times(I_{write}-I_{C_0})}{c+\log_{e}(\pi^2\times\Delta/4)\times (e\times m\times (1+p^2))})
	\end{split}
	\end{flalign}
where, \textit{P$_{write}$} is the probability of cell transition, \textit{I$_{write}$} is write current, \textit{c} is Euler constant, \textit{e} is electron charge, \textit{m}  is magnetic momentum of the free layer, \textit{p} is tunneling spin polarization, \textit{$\mu$$_{\beta}$} is Bohr magneton, and \textit{t$_{write}$} is write pulse width.

\begin{figure*}[btp]
				\centering\vspace{-15pt}
				\subfloat[]{\includegraphics[width=0.85\linewidth]{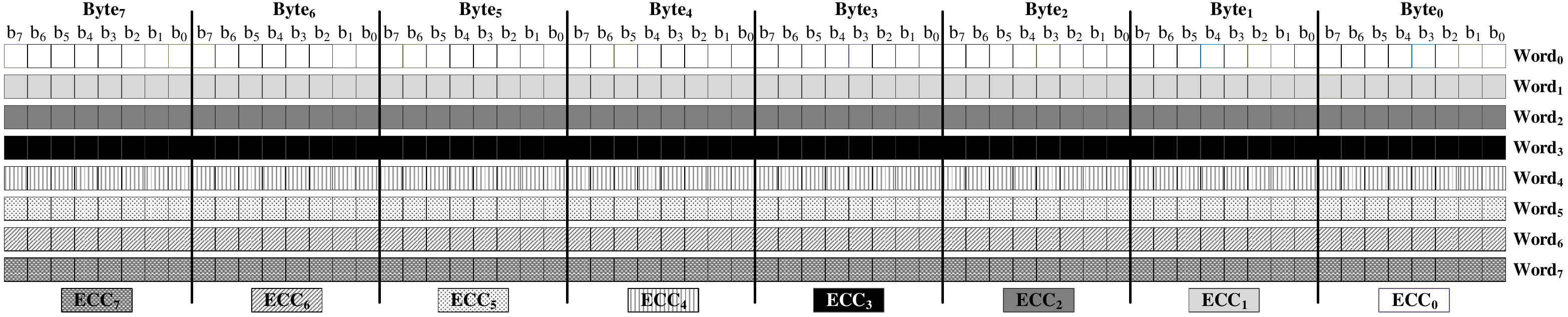}}\vspace{-5pt}
				\hspace{5pt}
				\subfloat[]{\includegraphics[width=0.85\linewidth]{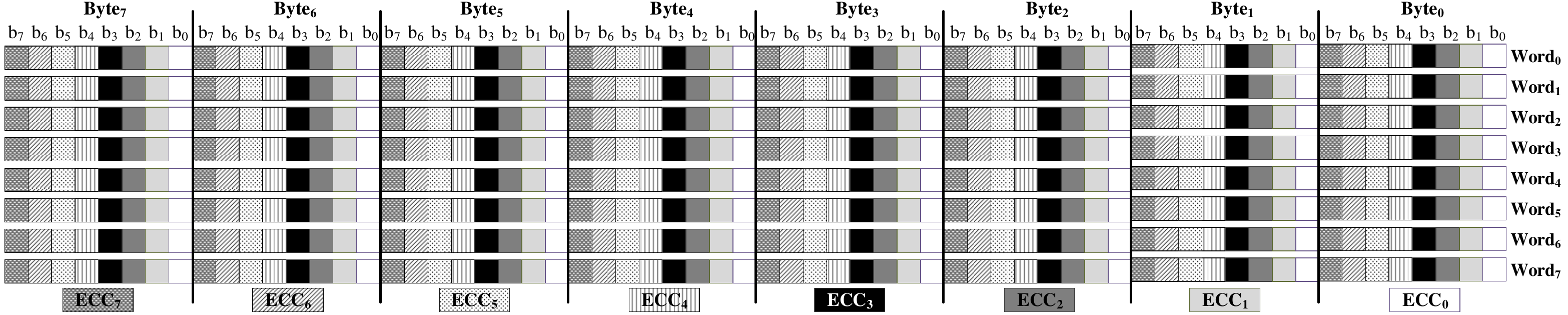}}\vspace{-10pt}
				\caption{ECC configuration schematic: (a) per-word configuration and (b) interleaved configuration.}\vspace{-20pt}
				\label{fig:2}
		\end{figure*}

To increase the probability of bit transition and reduce the write failure rate, previous studies increase the amplitude and/or width of the write pulse~\cite{sun-TMAG-12, lakys2012self}.
In addition to imposing high latency and energy overhead, these techniques increase the probability of oxide barrier breakdown in STT-MRAM cells. 
Another method 
is verifying each write operation by reading and comparing the written cell with incoming value and repeating the write operation on a mismatch~\cite{wrv}. 
Extra read operations increase the occurrence probability of read disturbance in this technique and higher dynamic energy is imposed due to extra write operations. 
Some studies try to reduce the number of bit transitions in write operations by encoding the incoming data or writing into a cache block with the minimum hamming distance~\cite{16-AZAD-TETC, AMir2016TDMR}. 
These techniques complicate the design and increase the read disturbance rate.

The most conventional and widely-used scheme to overcome write failures is to employ ECCs in cache blocks~\cite{ZAZADTPDS,farbeh2016floating,ZAZADTE, guo2017sanitizer, 16-AZAD-TETC}. 
ECC have been used in SRAM caches for decades and are the predominated protection scheme in modern processors.
This scheme is inherited by STT-MRAM caches and several studies utilized ECCs for overcoming write failure. For an ECC-protected STT-MRAM cache block, a write operation is successful as long as the number of erroneous cells due to write failure is not larger than the number of correctable errors.
\vspace{-10pt}

\section{Motivation and Proposed Method}

An ECC-protected LLC block is conventionally partitioned into \textit{M} codewords each consist of a \textit{k}-bit logical dataword and an \textit{r}-bit ECC generated by that dataword capable of correcting up to \textit{t} bits errors.
A codeword is written correctly if the number of write failures in its \textit{k+r} bits is less than or equal to \textit{t} and the write operation of a cache block is successful \textit{if and only if} all its codewords are written correctly. 
In this section, we formulate the probability of correct write operation for STT-MRAM caches protected by the existing ECC schemes and demonstrate their shortcomings in correcting write failure, based on our observations and investigations.


\vspace{-10pt}
\subsection{Problem Formulation} 
In today's processors, a cache block conventionally consists of eight 64-bit dataword each protected by an 8-bit \textit{Single Error Correction-Double Error Detection} (SEC-DED (72, 64)) code. For the sake of simplicity, we limit our discussion to the mentioned structure hereafter. However, the formulations, discussions, configurations, and proposed method are generally valid and applicable and not limited to this structure.

There are two configurations for generating eight SEC-DED(72, 64) codes in a cache block with 512 bits data. 
In the first configuration, known as \textit{per-word ECC}, eight consecutive data bytes are grouped to generate a SEC-DED code. 
In the second configuration, known as \textit{interleaved ECC}, bits with similar position in all 64 bytes contribute in generating each SEC-DED code.
Interleaved ECC is capable of correcting \textit{Multi-Bit Upsets} (MBUs) in SRAM caches by preventing the adjacent data bits to be grouped in the same logical dataword.
The configuration of per-word and interleaved SEC-DED(72, 64) is illustrated in Fig.~\ref{fig:2}(a) and Fig.~\ref{fig:2}(b), respectively.

Both per-word and interleaved ECCs are applicable to STT-MRAM LLCs and none of the recent studies has differentiated between these two configurations in term of error correction capability. 
The advantage of interleaved ECC over per-word ECC in SRAM LLCs for correcting MBUs is not valid for STT-MRAM LLCs, because write failure in the cells is independent of the cells adjacency. 
A codeword is written correctly if write failure occurs  in \textit{none} of its bits or \textit{only} in a single bit. 
Assuming 64-bit dataword protected by an 8-bit SEC-DED, the probability of a correct write operation is according to (\ref{eq:2}):
\vspace{-5pt}
	\begin{flalign}
	\begin{split}
			\label{eq:2}
			P_{word}= p_{write}^{k} + \binom k1 \times p_{write}^{(k-1)}(1-p_{write})
	\end{split}
	\end{flalign}
where, \textit{$P_{write}$} is the probability of successful transition of single bit (calculated in (\ref{eq:1})) and \textit{k} is the number of bits in the codeword that must be flipped.
All codewords must be written correctly in writing a cache block. The probability of a successful write operation for a block is according to (\ref{eq:3}):\vspace{-10pt}
	\begin{flalign}\vspace{-10pt}
	\begin{split}
			\label{eq:3}
			\shoveright{P_{block}= {P_{w}}_0 \times {P_{w}}_1 \times ...  \times {P_{w}}_7} 
			 \shoveright{=\prod_{i=1}^{7} {P_{w}}_i}\\ 
			  \shoveright{=\prod_{i=1}^{7}{[P_{write}^{k_{i}} + \binom {k_{i}}1 \times P_{write}^{(k_{i}-1)}(1-P_{write})}]} 
	\end{split}
	\end{flalign}
where, \textit{$w_i$} is codeword \textit{i} in the block and \textit{$k_i$} is the number of transitions in \textit{$w_i$}.

$P_{block}$ depends on the total number of transitions required for the target block as well as the distribution of these transitions between the codewords. The higher number of transitions and/or the larger variation in distribution of transitions, the lower probability of successful write operation is experienced. While the former depends on the data content and applications behavior, the latter depends on partitioning the data bits between codewords, which is directly determined by the ECC configuration. 

When the total number of required transitions in a block ($K = \sum_{i=1}^{7}{k_{i}}$) is fixed, it can be easily proven that the maximum value of \textit{$P_{block}$} in (\ref{eq:3}), which is the multiplication of probability of successful write of all codewords, is obtained when all \textit{$k_{i}$s} are the same and equal to $K/8$. 
Therefore, \textit{$P_{block}$} increases by more uniform  distribution of total transitions between the codewords and degrades by larger variance of that distribution.

\begin{table}[t]\vspace{-5pt}
				\centering
				\caption{Configuration of On-Chip Caches}\vspace{-10pt}
				\includegraphics[width=1\linewidth]{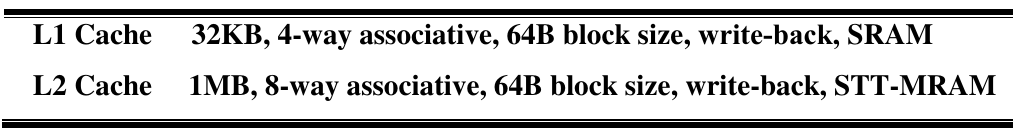}
				\label{table:1}\vspace{-25pt}
			\end{table}

\vspace{-2pt}
\subsection{Observation and Motivation}
\begin{figure}[t]
				\centering\vspace{-10pt}
				\subfloat[]{\includegraphics[width=0.87\linewidth]{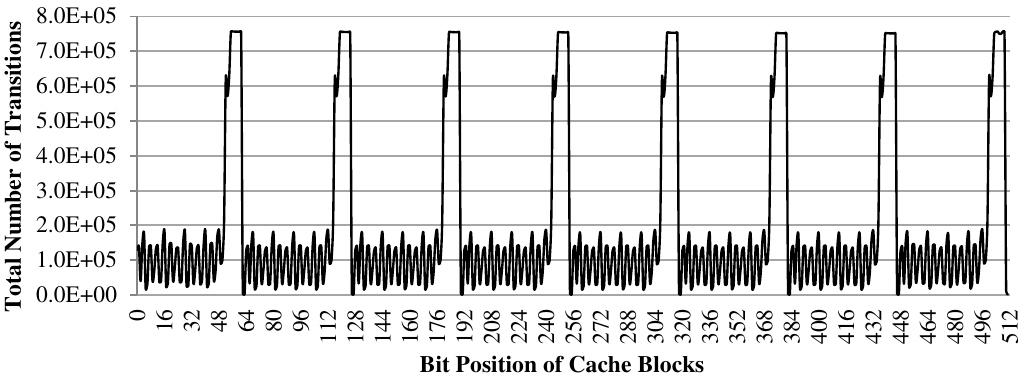}}\vspace{-2pt}
				\hspace{5pt}
				\subfloat[]{\includegraphics[width=0.87\linewidth]{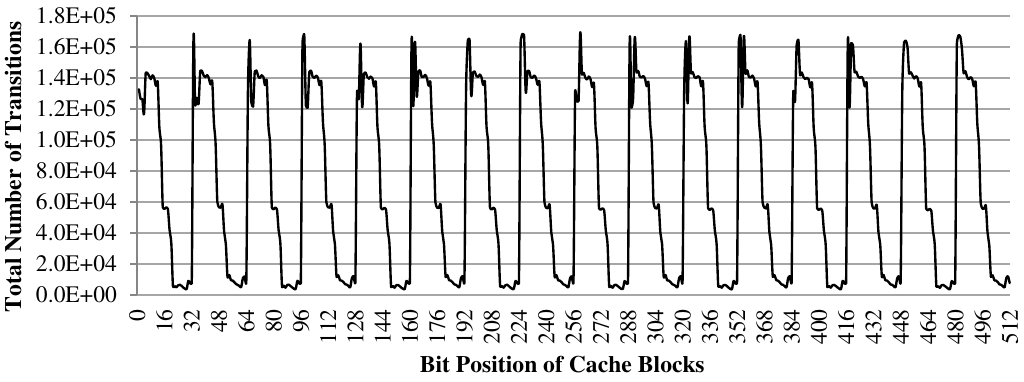}}\vspace{-2pt}
				\hspace{5pt}
				\subfloat[]{\includegraphics[width=0.87\linewidth]{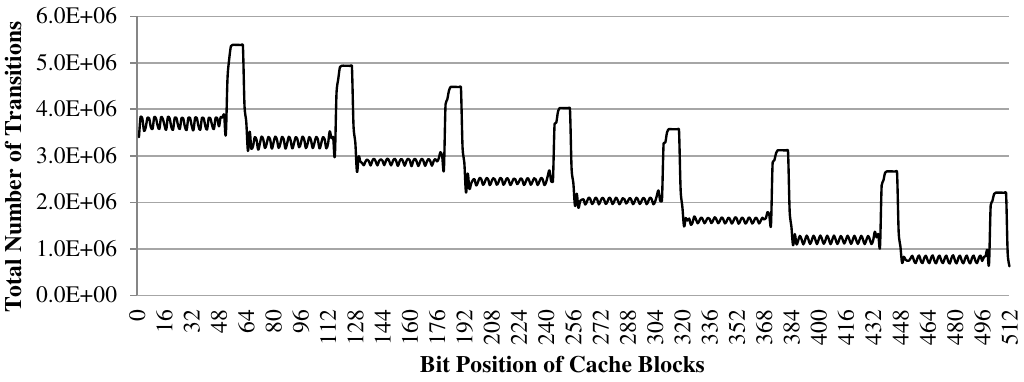}}\vspace{-2pt}
				\hspace{5pt}
				\subfloat[]{\includegraphics[width=0.87\linewidth]{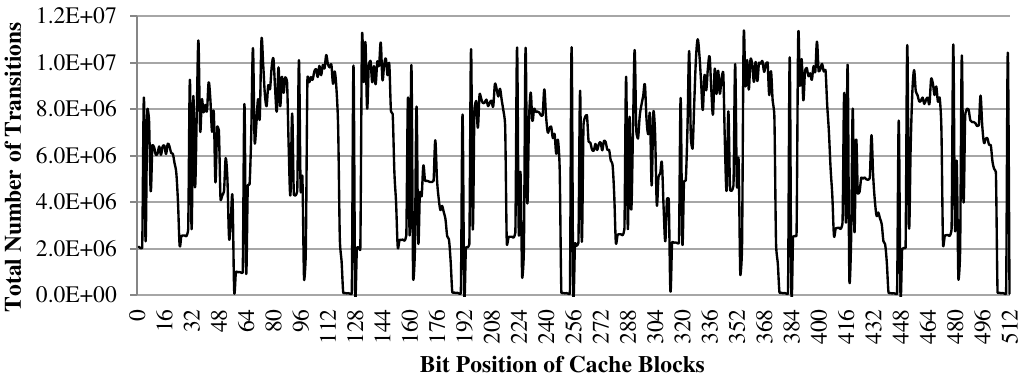}}\vspace{-10pt}
				\caption{Number of transitions in all bit positions of cache blocks: (a) bwaves, (b) calculix, (c) cactusADM, and (d) mcf workloads.}\vspace{-25pt}
				\label{fig:3}
			\end{figure}

\vspace{-1pt}
To investigate the capability of per-word and interleaved ECC configurations in evenly distributing the block transitions between their codewords, we conduct a set of simulations on SPEC CPU2006 benchmark suite using gem5 cycle-accurate simulator~\cite{gem5}.
The details of simulation is described in Table~\ref{table:1}.
Fig.~\ref{fig:3} shows the total number of transitions for write operations in all bit positions of cache blocks during the workload execution.
Considering the transitions of \textit{bwaves} workload in Fig.~\ref{fig:3}(a), the number of transitions in 20\% of upper part of each 64-bit word is by 4x larger than that of lower part of each word.
On the other hand, the transition pattern in all eight 64-bit words are almost similar.
For workloads with this write pattern, which is the case for \textit{floating-point} workloads, the distribution of transitions in per-word ECC can be more uniform than that in interleaved ECC.

Fig. \ref{fig:3}(b) depicts another write pattern in which similar behavior is observed for every 32-bit data.
In this pattern, which corresponds to \textit{calculix} as an \textit{integer} workload, the number of transitions sharply drops in such a way that the low part of data is more active than the upper part by 14x.
The write pattern of each 64-bit data in Fig.~\ref{fig:3}(c) is almost similar to that of Fig.~\ref{fig:3}(a), while the number of transitions in different words is largely different.
In this workload, the probability of existing a valid data in a block decreases by increasing the position number of words and a fraction of the block containing invalid data.
Considering \textit{mcf} workload, an irregular behavior is observed in the write pattern depicted in Fig. \ref{fig:3}(d).
In this pattern, the number of transitions is almost randomly distributed.
Even in this random pattern, a sharp drop can be observed in bit positions near 32, 64, 96, and so on.

The capability of both per-word and interleaved ECC configurations is not easy to predict and strongly depends on per write pattern.~An optimal ECC configuration, which provides the maximum probability of correct write operation, is the one that can evenly distribute the transitions between codewords in all write patterns.
To demonstrate how the non-uniformity in transition distribution degrades the cache reliability, we calculate the cache error rate for per-word and interleaved configurations and compare the results with error rate of optimal configuration.
Fig.~\ref{fig:4} depicts the normalized error rate for three mentioned ECC configurations.~The results show that the error rate in per-word and interleaved configurations is higher than that of optimal by 151.7\% and 42.3\%, respectively.~This value for per-word ECC is as high as 544.0\% in \textit{cactusADM} and as low as 10.5\% in \textit{astar} workload.~The worst-case error rate for interleaved ECC is 225.1\%, which is observed in $h264ref$ workload.

\subsection{Proposed Method}\vspace{-3pt}

\begin{figure}[t]
				\centering\vspace{-10pt}
				\includegraphics[width=0.9\linewidth]{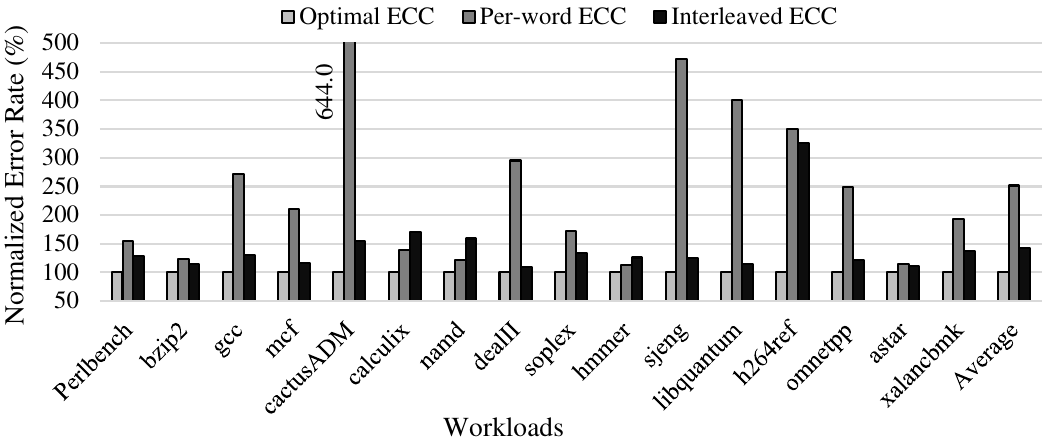}\vspace{-12pt}
				\caption{Cache error rate in per-word and interleaved ECC normalized optimal ECC.}\vspace{-26pt}
				\label{fig:4}
	      \end{figure}	

Write patterns vary in different applications and even in different write accesses per application. 
We have already observed that this variation leads to non-uniform distribution of bit transitions and inefficiency of both per-word and interleaved ECCs in correcting errors. 
An optimal ECC, which evenly distribute the total number of transitions between codewords in all write accesses minimizes the cache error rate and provides the maximum reliability. 
	
As observed in Fig.~\ref{fig:4}, 
for the majority of workloads, the error rate in per-word and interleaved configurations is significantly higher than that of optimal configuration.~An efficient ECC should provide a near-optimal error rate for all workloads and all write accesses within a workload.
Our proposed ECC configuration, so-called ${Inc\underline{r}emental~\underline{Ob}lique~\underline{In}terleaved}$~(ROBIN) ~ECC, generates the codewords in such a way that variation in workloads behaviors has minimum impact on the ECC efficiency and its error rate is not more than 10\% higher than that of the optimal ECC.

Prior to explain the architectural details of ROBIN, here we elaborate how various workload behaviors cause different transition distributions.~This discussion clarifies the key idea behind ROBIN and highlights its capability in uniformly distributing the transitions in various write patterns.~The transition patterns of write operations are data-dependent and beside the content, different data types cause different patterns.
As an example, in \textit{floating-point} applications, 
the number of transitions in exponent part is significantly larger than that in the mantissa part.~Our observations in Fig. \ref{fig:3}(a) and Fig. \ref{fig:3}(c) for \textit{bwaves} and \textit{cactusADM} workloads confirm this intuition. 

\begin{figure*}[h]
				\centering
				\includegraphics[width=0.9\linewidth]{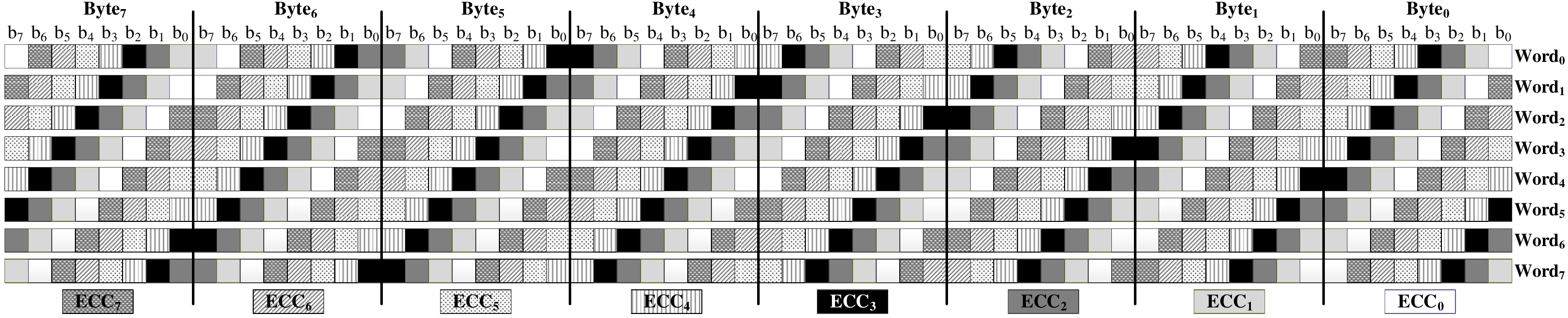}\vspace{-10pt}
				\caption{ECC configuration schematic for ROBIN ECC.}\vspace{-10pt}
				\label{fig:5}\vspace{-10pt}
	\end{figure*}
	
Considering \textit{integer} applications, cache blocks contain multiple 32-bit data words in which the transition probability in lower order bits is larger than that in higher order bits.
Majority of integer data are \textit{narrow-width values}~\cite{imani2016low}, which confirm our intuition on larger number of transitions in lower order bits. 
We have already observed this behavior in Fig. \ref{fig:3}(b) for $calculix$ workload.
Applications with large fraction of string data type have their unique write pattern, which strongly depends on the character coding standards.
Besides the type of data, the validity of cache block words affects the transition pattern.
There can be blocks in which all their words contain valid data, e.g., $bwaves$ workload in Fig. \ref{fig:3}(a), and blocks with several invalid words, e.g., some written blocks for $cactusADM$ as shown in Fig. \ref{fig:3}(c).

An efficient ECC configuration should be capable of evenly distributing the total transitions between all codewords.~To this aim, ROBIN partitions the data bits in such a way that all 64-bit words in a block as well as all eight bytes of these words contribute in generating all ECC words.~In addition, all bit positions of all bytes within a word contribute similarly in each codeword.

Fig.~\ref{fig:5} illustrates the ROBIN policy in constructing codewords and generating ECCs.
Considering a 512-bit cache block consisting eight 64-bit data words, eight bits from each data word contribute in generating each ECC.
From another aspect, all 64 bytes contribute in generating all ECCs (one bit from each byte for each ECC).
In addition, eight bit positions in each eight bytes of the word are similarly distributed between the ECCs.
For $byte_0$ in the words, all bit positions $b_0$, $b_1$, ..., $b_7$ each selected from $byte_0$ of a unique word are in the same group.~To balance the variation between bit positions of different bytes inside a word, a unique bit position from each byte of a word is selected for generating each ECC.
Mathematically speaking, each ECC word $ECC_n$ (n $\in$ \{0, ,1, ...,7\}) is a function of 64 data bits selected according to~(\ref{eq:4}):
	\begin{flalign}\vspace{-20pt}
	\begin{split}
			\label{eq:4}
			ECC_{n}= f(\{b_{i,j,(i+j+n)} |~\forall i,j  \{0, 1, ..., 7\}\})
	\end{split}
	\end{flalign}
where, $b_{i,j,(i+j+n)}$ is a data bit in position $(i+j+n)mod(8)$ in $byte_j$ of $word_i$.
Using this policy in grouping data bits to generate ECCs, ROBIN is capable of distributing transition variations between data words, between bytes of each word, and between bit positions of each byte.
For the sake of clarity, we focused on 512-bit cache blocks protected by eight 8-bit SEC-DED codes.
However, ROBIN configuration is generally independent of data and ECC size as well as coding scheme.
It is scalable in term of block size and is also applicable to other ECCs.
This method has no effect on the complexity of ECC encoder/decoder logic and its overhead is almost the same as those of interleaved ECC.
 \begin{figure*}[t]
				\centering
				\includegraphics[width=0.9\linewidth]{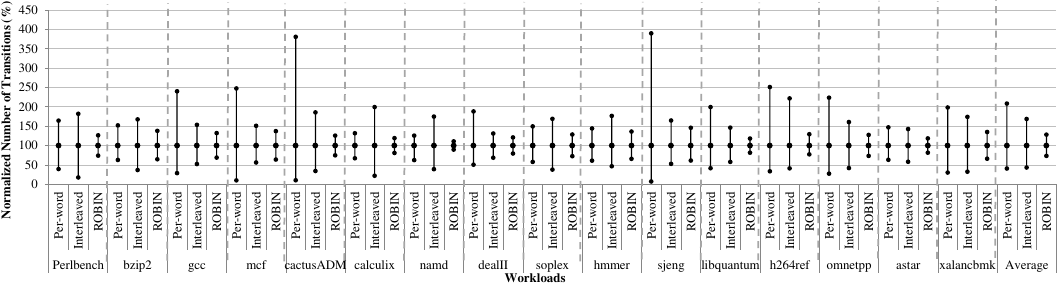}\vspace{-5pt}
				\caption{Variation in number of transitions between codewords of cache blocks (minimum, average, and maximum).}\vspace{-20pt}
				\label{fig:6}
	      \end{figure*}

\vspace{-10pt}
\section{Simulation Setup and Results}\vspace{-3pt}

To evaluate the efficiency of the proposed ROBIN ECC, we implement it in gem5 cycle-accurate simulator~\cite{gem5} and use SPEC CPU2006 benchmark suite as workloads.
The first 100 million instructions are skipped as warm-up phase and the results are extracted for the next one billion instructions.
The details of simulation was shown in Table~\ref{table:1}.
ROBIN is compared with per-word and interleaved ECC in terms of transitions distribution and error rates.
The results are normalized to optimal ECC.

Fig.~\ref{fig:6} depicts the variation in the total number of transitions in cache blocks.
For all write accesses in a workload, we sort the number of transitions in eight codewords in a block and the summation of these values are normalized to the average of transitions in codewords (which is the number of transitions in optimal ECC).
In Fig.~\ref{fig:6}, the number of transitions in codewords with minimum and maximum transitions is depicted for per-word, interleaved, and ROBIN configurations.

Normalizing to the number of transitions in a codeword for uniform distribution (optimal), the results show that the number of transitions in codewords of per-word ECC is from 41.1\% to 208.8\%, on average.
This interval is smaller for interleaved ECC and the minimum and maximum transitions in codewords are 43.6\% and 169.2\%, respectively. 
ROBIN significantly reduces this gap by limiting the number of transitions in codewords between 73.7\% to 128.4\%, on average. 
The interesting achievement is that in the worst case, the minimum and maximum transitions of codeword in ROBIN is larger than 61.8\% and lower than 146.1\%, respectively.
These values for per-word ECC are 7.4\% and 390.4\%, respectively, and for interleaved ECC are 17.8\% and 222.2\%, respectively.  
The wider gap between the number of transitions in codewords indicates larger non-uniformity, which results in higher cache error rate compared to uniform distribution (optimal ECC).

 \begin{figure}[t]
				\centering
				\includegraphics[width=0.9\linewidth]{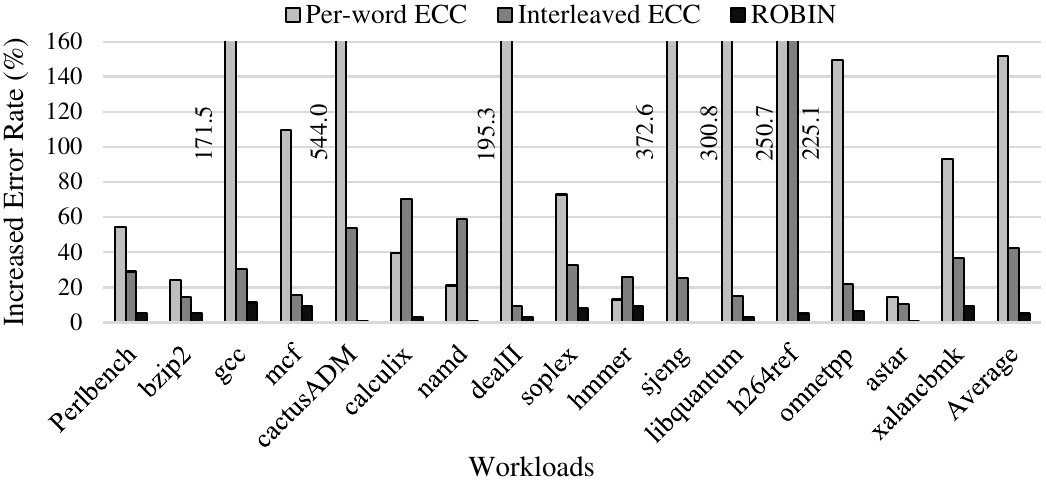}\vspace{-10pt}
				\caption{Cache error rate increase for per-word, interleaved, and ROBIN ECC configurations normalized to the optimal ECC.}\vspace{-25pt}
				\label{fig:7}
	      \end{figure}

Fig.~\ref{fig:7} depicts the cache error rate in three evaluated ECC configurations.~The results illustrate the increase in cache error rate due to non-uniformity of transitions distributions compared to the optimal ECC.
On average, per-word and interleaved ECC increases the error rate by 151.7\% and 42.3\%, respectively.
This value is reduced to 5.3\% in ROBIN.
Therefore, ROBIN reduces the error rate for per-word and interleaved ECC due to non-uniformity of transitions by 28.6x and 8.0x, respectively.
For some workloads, error rate in per-word increases by even more than 300\%, e.g.,~$cactusADM$,~$sjeng$,~and~$libquantum$.
In the worst-case, interleaved ECC increases the error rate by 225.1\% in $h264ref$.
The error rate increase in ROBIN is less than 10\% for all workloads.
This value is even less than 1\% for some workloads, e.g.,~$cactusADM$, $namd$, $sjeng$, and $astar$, which is almost the same as optimal ECC.


\vspace{-15pt}
\section{Summary and Conclusions}\vspace{-3pt}
Write failure is the main reliability challenge in emerging STT-MRAM caches.
Data dependency and variations in different codewords of a block significantly degrade the efficiency of conventional ECCs.
This paper 1) investigated the commonly used per-word and interleaved ECC and revealed that these schemes increase the cache error rate by 151.7\% and 42.3\%, respectively, and
2) proposed the efficient ROBIN ECC to reduce the cache error rate to as low as 5.3\%.
This significant improvement in cache reliability is achieved with no increase in ECC overheads.
ROBIN is a promising alternative for conventional per-word and interleaved configurations to increase the efficiency of ECCs in emerging STT-MRAM caches.
\vspace{-10pt}
\section*{Acknowledgment}\vspace{-9pt}
\begin{spacing}{1.2}
This work has been partially supported by $Iran~National~Science~Foundation$ (INSF) under grant number 96006071 and $National~Elites~Foundation$.
\end{spacing}
\vspace{-10pt}
\def\bibfont{\footnotesize}\footnotesize       
\bibliographystyle{IEEEtran}
\bibliography{IEEEabrv,references}

@article{Eli-TC,
  title={TA-LRW: a replacement policy for error rate reduction in STT-MRAM caches},
  author={Cheshmikhani, Elham and Farbeh, Hamed and Miremadi, Seyed Ghassem and Asadi, Hossein},
  journal={IEEE Trans. Comput.},
   year={in press, 2018}
}

@article{guo2017sanitizer,
  title={Sanitizer: mitigating the impact of expensive ecc checks on stt-mram based main memories},
  author={Guo, Xiaochen and Bojnordi, Mahdi Nazm and Guo, Qing and Ipek, Engin},
  journal={IEEE Trans. Comput.},
  year={2017, in press},
}

@article{farbeh2016floating,
  title={Floating-ECC: Dynamic repositioning of error correcting code bits for extending the lifetime of STT-RAMs},
  author={Farbeh, Hamed and Kim, Hyeonggyu and Miremadi, Seyed Ghassem and Kim, Soontae},
  journal={IEEE Trans. Comput.},
  volume={65},
  number={12},
  pages={3661--3675},
  year={2016},
  publisher={IEEE}
}

@article{ZAZADTE,
  title={{AWARE: adaptive way allocation for reconfigurable ECCs to protect write errors in STT-RAM caches}},
  author={Azad, Zahra and Farbeh, Hamed and Monazzah, Amir Mahdi Hosseini and Miremadi, Seyed Ghassem},
  journal={IEEE Trans. Emerging Topics Comput.},
  year={2017, in press}
}

@article{Chen2016,
  title={{A review of emerging non-volatile memory (NVM) technologies and applications}},
  author={Chen, An},
  journal={Elsevier Solid-State Electron.},
  volume={125},
  pages={25--38},  
  year={2016}
}

@article{choi2017nvm,
  title={NVM way allocation scheme to reduce NVM writes for hybrid cache architecture in chip-multiprocessors},
  author={Choi, Juhee and Park, Gi-Ho},
  journal={IEEE Trans. Paral. Dist. Syst.},
  volume={28},
  number={10},
  pages={2896--2910},  
  year={2017}
}

@article{sun-TMAG-12,
  title={Architectural exploration to enable sufficient mtj device write margin for stt-ram based cache},
  author={Sun, Hongbin and others},
  journal={IEEE Trans. Mag.},
  volume={48},
  number={8},
  pages={2346--2351},
  year={2012}
}

@inproceedings{16-AZAD-TETC,
  title={CD-ECC: Content-Dependent Error Correction Codes for combating asymmetric nonvolatile memory operation errors},
  author={Wen, Wujie and others},
  booktitle={Proc. Int. Conf. Comput.-Aided Des.},
  year={2013}
  }

@article{vatajelu2017challenges,
  title={Challenges and solutions in emerging memory testing},
  author={Vatajelu, Elena Ioana and others},
  journal={IEEE Trans. Emerging Topics Comput.},
  year={2017, in press}
}

@article{mittal2017survey,
  title={A survey of soft-error mitigation techniques for non-volatile memories},
  author={Mittal, Sparsh},
  journal={Computers},
  volume={6},
  number={1},
  pages={8},
  year={2017},
  publisher={Multidisciplinary Digital Publishing Institute}
}

@inproceedings{imani2016low,
  title={A low-power hybrid magnetic cache architecture exploiting narrow-width values},
  author={Imani, Mohsen and Rahimi, Abbas and Kim, Yeseong and Rosing, Tajana},
  booktitle={Proc IEEE Non-Vol. Mem. Syst. App. Symp.},
  year={2016}
}

@inproceedings{asadi2017wipe,
  title={WIPE: Wearout informed pattern elimination to improve the endurance of nvm-based caches},
  author={Asadi, Sina and Monazzah, Amir Mahdi Hosseini and Farbeh, Hamed and Miremadi, Seyed Ghassem},
  booktitle={Proc. IEEE Asia South Pacific Des. Automat. Conf.},
  year={2017}
}

@article{gem5,
  title={{The gem5 simulator}},
  author={Binkert, Nathan and others},
  journal={ACM SIGARCH Comput. Arch. News},
  volume={39},
  number={2},
  pages={1--7},
  year={2011}
}

@article{wrv,
  title={{Improving reliability of non-volatile memory technologies through circuit level techniques and error control coding}},
  author={Yang, Chengen},
  journal={EURASIP J.Adv. Sig. Proc.},
  pages={1--24},
  year={2012}
}

@article{ZAZADTPDS,
  title={An efficient protection technique for last level STT-RAM caches in multi-core processors},
  author={Azad, Zahra and Farbeh, Hamed and Monazzah, Amir Mahdi Hosseini and Miremadi, Seyed Ghassem},
  journal={IEEE Trans. Paral. Dist. Syst.},
  volume={28},
  number={6},
  pages={1564--1577},
  year={2017}
}

@article{14-zazad-eken2014novel,
  title={A novel self-reference technique for stt-ram read and write reliability enhancement},
  author={Eken, Enes and others},
  journal={IEEE Trans. Mag.},
  volume={50},
  number={11},
  pages={1--4},
  year={2014}
}

@article{naeimi2013intel,
  title={STT-MRAM scaling and retention failure},
  author={Naeimi, Helia and others},
  journal={Intel Tech. J.},
  volume={17},
  number={1},
  pages={54--75},
  year={2013}
}

@article{Chintaluri-ESTCS2016,
 title={Analysis of defects and variations in embedded Spin Transfer Torque (STT) MRAM arrays},
  author={Chintaluri, Ashwin and Naeimi, Helia and Natarajan, Suriyaprakash and Raychowdhury, Arijit},
  journal={IEEE J. Emerg. Sel. Top. Circ. Syst.},
  volume={6},
  number={3},
  pages={319--329},
  year={2016}
}

@article{lakys2012self,
  title={Self-enabled error-free switching circuit for spin transfer torque MRAM and logic},
  author={Lakys, Yahya and others},
  journal={IEEE Trans. Mag.},
  volume={48},
  number={9},
  pages={2403--2406},
  year={2012},
  publisher={IEEE}
}

@article{AMir2016TDMR,
  title={{Ler: Least error rate replacement algorithm for emerging STT-RAM caches}},
  author={Monazzah, Amir Mahdi Hosseini and Farbeh, Hamed and Miremadi, Seyyed Ghassem},
  journal={IEEE Trans. Device Mater. Rel.},
  volume={16},
  number={2},
  pages={220--226},
  year={2016}
}

@inproceedings{salkhordeh2016operating,
  title={An operating system level data migration scheme in hybrid DRAM-NVM memory architecture},
  author={Salkhordeh, Reza and Asadi, Hossein},
  booktitle={Proc. Des., Autom. \& Test Euro. Conf. \& Exh.},
  pages={936--941},
  year={2016},
  organization={EDA Consortium}
}
%
%
%
%
\end{document}